\documentclass[pdftex]{algotel}

\usepackage[latin1]{inputenc}
\usepackage[francais]{babel}
\usepackage{float}
\usepackage{amsmath}
\usepackage{graphicx}
\usepackage{amssymb}
\usepackage{theorem}
\usepackage{float}
\usepackage{verbatim}
\usepackage{wrapfigure}

\floatstyle{ruled}
\newfloat{algorithm}{tbp}{loa}
\floatname{algorithm}{Algorithme}

\newtheorem{theo}{Théorème}
\newtheorem{defi}{Définition}
\newtheorem{lem}{Lemme}
\newtheorem{spec}{Spécification}
\newenvironment{theorem}[1]{\vspace{-0.35cm}\begin{theo}#1}{\end{theo}\vspace{-0.3cm}}
\newenvironment{definition}[1]{\vspace{-0.35cm}\begin{defi}#1}{\end{defi}\vspace{-0.3cm}}

\newenvironment{specification}[1]{\vspace{-0.35cm}\begin{spec}#1}{\end{spec}\vspace{-0.3cm}}

\author{Swan Dubois et Rachid Guerraoui}
\address{\'Ecole Polytechnique Fédérale de Lausanne (Suisse), \{swan.dubois,rachid.guerraoui\}@epfl.ch}

\title[Spéculation et auto-stabilisation]{Spéculation et auto-stabilisation}

\keywords{Spéculation, Tolérance aux fautes, Auto-stabilisation, Exclusion mutuelle.}

\begin{document}
\maketitle

\begin{abstract}
L'auto-stabilisation garantit qu'à la fin d'une période de fautes transitoires, un système réparti retrouve de lui-même un comportement correct en un temps fini. La spéculation consiste à garantir que le système soit correct pour toute exécution mais possède des performances significativement meilleures pour un sous-ensemble d'exécutions qui sont supposées plus probables. Un protocole spéculatif est donc à la fois robuste et efficace en pratique. Nous introduisons ici la notion de spéculation en auto-stabilisation en présentant un protocole spéculativement stabilisant d'exclusion mutuelle. Ce dernier stabilise pour toute exécution et son temps de stabilisation est optimal pour les exécutions synchrones.
\end{abstract}

\section{Motivations}

L'approche spéculative \cite{L08c} repose sur l'existence d'un compromis permanent entre la robustesse et l'efficacité des protocoles répartis. En effet, il est demandé aux applications réparties de tolérer à la fois un grand nombre de conditions difficiles (répartition des données, asynchronisme, fautes...) ainsi que de fournir les meilleures performances possibles (principalement en temps). Cependant, garantir la robustesse du protocole repose sur des mécanismes comme la synchronisation ou la réplication qui induisent généralement de mauvaises performances. L'approche spéculative suppose alors que, même si des exécutions présentant de mauvaises conditions sont toujours possibles, certaines exécutions favorables (par exemple synchrones et sans fautes) sont plus probables. L'idée est alors de garantir que le protocole restera correct quelles que soient les conditions de l'exécution mais sera optimisé pour un sous-ensemble d'exécutions qui sont les plus probables en pratique. L'objectif de cet article est d'exploiter cette approche en auto-stabilisation.

L'auto-stabilisation \cite{D74j} est une technique de tolérance aux fautes transitoires (\emph{i.e.} de durée finie). Un système auto-stabilisant garantit qu'à la fin d'une faute transitoire (qui peut corrompre de manière arbitraire l'état du système), il retrouvera un comportement correct en un temps fini et sans aide extérieure. Dans cet article, nous définissons une nouvelle variante de l'auto-stabilisation dans laquelle la mesure principale de performance, le temps de stabilisation, est vue comme une fonction de l'adversaire et non comme une valeur unique. Nous associons à chaque adversaire (connu également sous le nom d'ordonnanceur ou de démon) le pire temps de stabilisation du protocole sur l'ensemble des exécutions décrites par cet adversaire. Nous pouvons alors définir un protocole spéculativement stabilisant  comme un protocole auto-stabilisant sous un adversaire donné mais qui présente un temps de stabilisation significativement meilleur sous un autre adversaire (plus faible). De cette manière, nous nous assurons que le protocole stabilise sur un large ensemble d'exécutions mais est efficace sur un ensemble d'exécutions plus restreint (mais plus probables).

Bien que cette notion de spéculation soit nouvelle dans le domaine de l'auto-stabilisation, certains protocoles existants vérifient notre définition, en quelque sorte par accident. Par exemple, la complexité du protocole d'exclusion mutuelle de Dijkstra \cite{D74j} tombe en $n$ étapes sous le démon synchrone (où $n$ est le nombre de processeurs). Cependant, ce résultat n'est pas optimal. La contribution principale de cet article est un nouveau protocole d'exclusion mutuelle spéculativement stabilisant. Nous prouvons que son temps de stabilisation pour les exécutions synchrones est de $\left\lceil diam(g)/2\right\rceil$ étapes (où $diam(g)$ est le diamètre du système), ce qui améliore significativement la borne du protocole de Dijkstra. En réalité, nous prouvons que cela est optimal car nous présentons un résultat de borne inférieure sur le temps de stabilisation de l'exclusion mutuelle pour les exécutions synchrones. Ce résultat est intéressant en lui-même étant donné qu'il est indépendant de la spéculation. Pour finir, notre protocole ne requiert aucune hypothèse sur la topologie du système contrairement à celui de Dijkstra.

\section{Modèle et définitions}

Nous considérons un système réparti, \emph{i.e.} un graphe non orienté connexe $g$ où les sommets représentent les processeurs et les arêtes représentent les liens de communication.  Deux processeurs $u$ et $v$ sont \emph{voisins} si l'arête $(u,v)$ existe dans $g$. L'ensemble des voisins de $v$ est noté $vois(v)$. Le nombre de processeurs et le diamètre du système sont respectivement notés $n$ et $diam(g)$. Chaque processeur $v$ possède une identité unique $id_v\in ID$. Nous supposons que $ID=\{0,\ldots,n-1\}$. Les variables d'un processeur définissent son \emph{état}. L'ensemble des états des processeurs du système à un instant donné forme la \emph{configuration} du système. L'ensemble des configurations du système est noté $\Gamma$. Nous prenons comme modèle de calcul le \emph{modèle à états}. Les variables des processeurs sont partagées : chaque processeur a un accès direct en lecture aux variables de ses voisins. En une \emph{étape} atomique, chaque processeur peut lire son état et ceux de ses voisins et modifier son propre état. Un \emph{protocole} est constitué d'un ensemble de règles de la forme $<garde>\longrightarrow<action>$. La $garde$ est un prédicat sur l'état du processeur et de ses voisins tandis que l'$action$ est une séquence d'instructions modifiant l'état du processeur. \`A chaque étape, chaque processeur évalue ses gardes. Il est dit \emph{activable} si l'une d'elles est vraie. Il est alors autorisé à exécuter son $action$ correspondante (en cas d'exécution simultanée, tous les processeurs activés prennent en compte l'état du système du début de l'étape). Les \emph{exécutions} du système (séquences d'étapes) sont gérées par un \emph{ordonnanceur} (ou \emph{démon}) : à chaque étape, il sélectionne au moins un processeur activable pour que celui-ci exécute sa règle. Cet ordonnanceur permet de modéliser l'asynchronisme du système. Il existe de nombreuses variantes de démons (\emph{cf.} \cite{DT11r}). Dans cet article, nous utiliserons le démon synchrone (à chaque étape, l'ensemble des processeurs activables sont sélectionnés par le démon), noté $ds$, et le démon inéquitable distribué (aucune contrainte n'est donnée au démon), noté $did$. Nous définissons l'ordre partiel suivant sur l'ensemble des démons : $d'\preccurlyeq d$ si l'ensemble des exécutions autorisées par $d'$ est inclus dans celui des exécutions autorisées par $d$. Le démon $d'$ est alors dit plus faible que $d$.

\begin{definition}[Auto-stabilisation \cite{D74j}]
Un protocole réparti $\pi$ est auto-stabilisant pour la spécification $spec$ sous un démon $d$ si, partant de toute configuration de $\Gamma$, toute exécution de $\pi$ sous $d$ contient une configuration à partir de laquelle toute exécution de $\pi$ sous $d$ vérifie $spec$. Nous notons $temps\_stab(\pi,d)$ le temps de stabilisation de $\pi$ sous $d$.
\end{definition}

Nous pouvons à présent introduire la définition principale de cet article qui formalise la notion de spéculation en auto-stabilisation.

\begin{definition}[Stabilisation spéculative]
Pour deux démons $d$ et $d'$ vérifiant $d'\prec d$, un protocole réparti $\pi$ est $(d,d',f)$-spéculativement stabilisant pour la spécification $spec$ si : $(i)$ $\pi$ est auto-stabilisant pour $spec$ sous $d$ et $(ii)$ $f$ est une fonction telle que : $temps\_stab(\pi,d)/temps\_stab(\pi,d')\in\Omega(f)$.
\end{definition}

\section{Exclusion mutuelle}

L'exclusion mutuelle est un problème fondamental qui consiste à assurer que tout processeur peut exécuter infiniment souvent une section particulière de son code, appelée section critique, avec la garantie qu'il n'y ait jamais deux processeurs qui exécutent simultanément leur section critique. Notre contribution sur ce problème est de présenter un nouveau protocole auto-stabilisant sous le démon inéquitable distribué qui présente un temps de stabilisation optimal sous le démon synchrone.

Nous adoptons la spécification suivante de l'exclusion mutuelle. Pour chaque processeur $v$, nous définis- sons un prédicat $privilege_v$. Un processeur $v$ est privilégié dans une configuration $\gamma$ si et seulement si $privilege_v=vrai$ dans $\gamma$. Si un processeur $v$ est privilégié dans une configuration $\gamma$ et que $v$ est activé durant l'étape $(\gamma,\gamma')$, alors $v$ exécute sa section critique durant cette étape.

\begin{specification}[Exclusion mutuelle $spec_{EM}$]
Une exécution $e$ vérifie $spec_{EM}$ si au plus un processeur est privilégié dans toute configuration de $e$ (sûreté) et si tout processeur exécute infiniment souvent sa section critique dans $e$ (vivacité).
\end{specification}

Notre protocole est basé sur un protocole d'unisson auto-stabilisant \cite{BPV04c}. Ce problème consiste  à assurer, sous le démon inéquitable distribué, des garanties sur les horloges logiques des processeurs. Chaque processeur possède un registre qui stocke la valeur actuelle de son horloge logique. Un protocole d'unisson assure alors que la différence entre les horloges de processeurs voisins est toujours bornée et que chaque horloge est infiniment souvent incrémentée. Dans la suite, nous résumons les résultats de \cite{BPV04c}.

\paragraph{Unisson.} Une horloge bornée $\mathcal{X}=(H,\phi)$ est un ensemble fini $H=cerise(\alpha,K)$ (paramétré par deux entiers $\alpha\geq 1$ et $K\geq 2$) doté d'une fonction d'incrémentation $\phi$ définie comme suit.
\begin{wrapfigure}{r}{6cm}
\noindent\begin{center} 
\includegraphics[height=5cm]{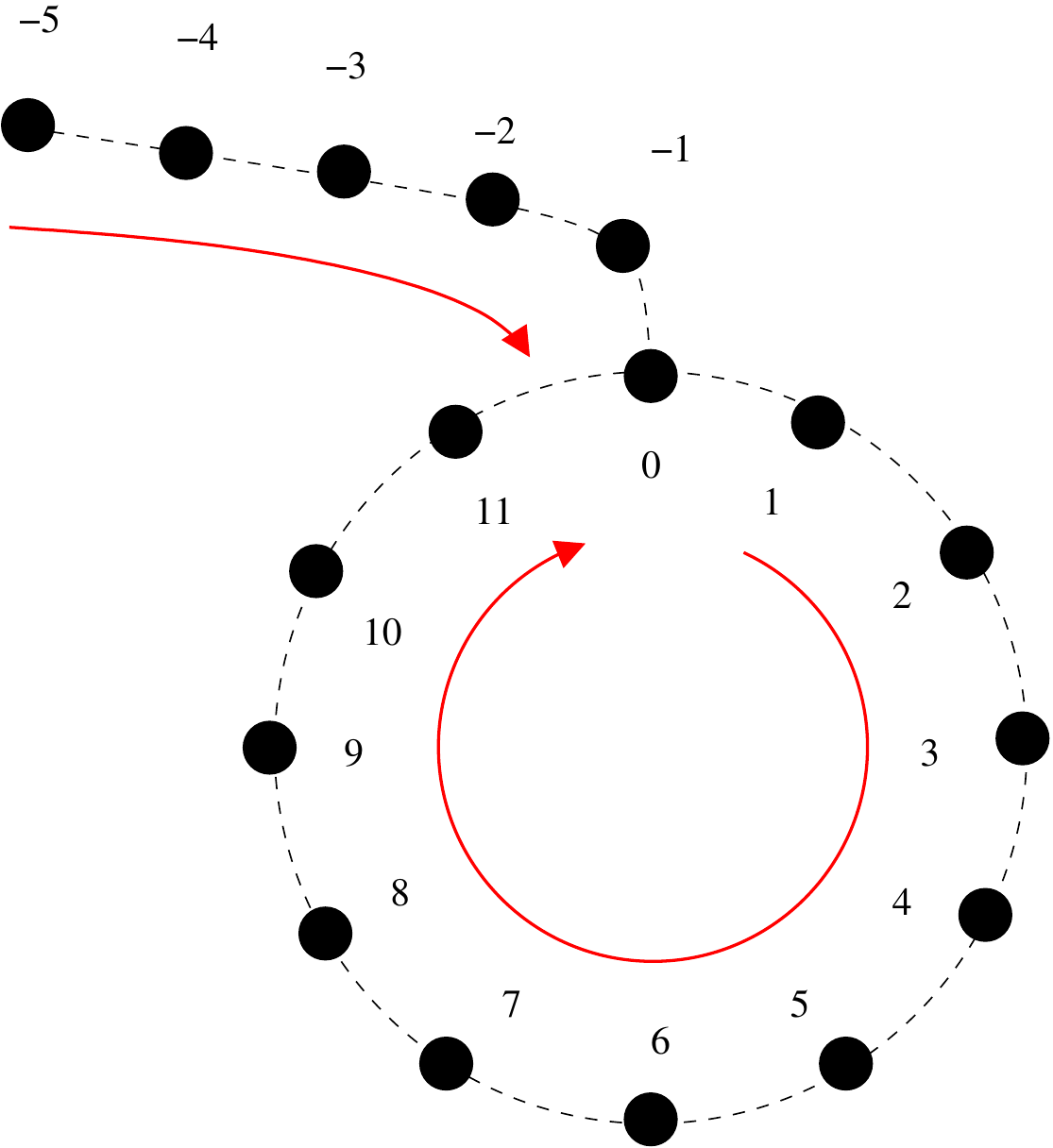}
\end{center}
\caption{Une horloge bornée $\mathcal{X}=(cerise(\alpha,K),\phi)$ avec $\alpha=5$ et $K=12$}
\label{fig:cherry}
\end{wrapfigure}
Soit $c$ un entier. Notons $\overline{c}$ l'unique élément de $[0,\ldots,K-1]$ tel que $c=\overline{c}$ mod $K$. Nous définissons la distance $d_K(c,c')=min\{\overline{c-c'},\overline{c'-c}\}$ sur $[0,\ldots,K-1]$. Deux entiers $c$ et $c'$ sont localement comparables si $d_K(a,b)\leq 1$. Nous définissons alors la relation d'ordre local $\leq_l$ comme suit : $c\leq_l c'$ si et seulement si $0\leq \overline{c'-c}\leq 1$. Définissons $cerise(\alpha,K)=\{-\alpha,\ldots,0,\ldots,K-1\}$. Soit $\phi$ la fonction définie par :
\[\phi:c\in cerise(\alpha,K)\mapsto\left\{\begin{array}{ll}
(c+1) & \text{si } c<0\\
(c+1) \text{ mod } K & \text{sinon}
\end{array}\right.\]

La paire $\mathcal{X}=(cerise(\alpha,K),\phi)$ est une horloge bornée de valeur initiale $-\alpha$ et de taille $K$ (voir Figure \ref{fig:cherry}). Une valeur d'horloge $c\in cerise(\alpha,K)$ est incrémentée quand cette valeur est remplacée par $\phi(c)$. Une ré-initialisation de $\mathcal{X}$ est une opération consistant à remplacer toute valeur de $cerise(\alpha,K)\setminus\{-\alpha\}$ par $-\alpha$. Soient respectivement $init_\mathcal{X}=\{-\alpha,\ldots,0\}$ et $stab_\mathcal{X}=\{0,\ldots,K-1\}$ les ensembles de valeurs initiales et correctes de $\mathcal{X}$. Nous notons $init^*_\mathcal{X}=init_\mathcal{X}\setminus\{0\}$, $stab^*_\mathcal{X}=stab_\mathcal{X}\setminus\{0\}$ et $\leq_{init}$ l'ordre total naturel sur $init_\mathcal{X}$.

Soit un système réparti dans lequel tout processeur $v$ a un registre $r_v$ stockant une valeur d'une horloge bornée $\mathcal{X}=(H,\phi)$ avec $H=cerise(\alpha,K)$. Nous définissons une configuration légitime pour l'unisson comme une configuration dans laquelle $\forall v\in V,\forall u\in vois(v),(r_v\in stab_\mathcal{X})\wedge(r_u\in stab_\mathcal{X})\wedge(d_K(r_v,r_u)\leq 1)$. En d'autres termes, une configuration légitime est une configuration telle que toute valeur d'horloge est correcte et l'écart entre les valeurs d'horloges de processeurs voisins est borné par $1$. Nous notons $\Gamma_1$ l'ensemble des configurations légitimes pour l'unisson. Il est important de noter que l'on a, pour toute configuration de $\Gamma_1$ et toute paire de processeurs $(u,v)$, $d_K(r_u,r_v)\leq diam(g)$.

\begin{specification}[Unisson $spec_{UA}$]
Une exécution $e$ vérifie $spec_{UA}$ si toute configuration de $e$ appartient à $\Gamma_1$ (sûreté) et que l'horloge de tout processeur est infiniment souvent incrémentée dans $e$ (vivacité).
\end{specification}

Dans \cite{BPV04c}, les auteurs proposent un protocole d'unisson auto-stabilisant sous le démon inéquitable distribué. L'idée principale est de ré-initialiser l'horloge de tout processeur qui détecte une violation locale de la condition de sûreté (\emph{i.e.} l'existence d'un voisin ayant une valeur d'horloge non localement comparable). Autrement, un processeur est autorisé à incrémenter son horloge (que sa valeur soit correcte ou initiale) seulement si cette dernière a la valeur minimale localement. Le choix des paramètres $\alpha$ et $K$ est crucial. En particulier, pour rendre le protocole auto-stabilisant sous le démon inéquitable distribué, ces paramètres doivent satisfaire $\alpha\geq trou(g)-2$ et $K>cyclo(g)$, où $trou(g)$ et $cyclo(g)$ sont deux constantes liées à la topologie de $g$. Plus précisément, $trou(g)$ est la taille du plus grand trou de $g$ (\emph{i.e.} du plus long cycle sans corde), si $g$ contient un cycle, $2$ sinon. $cyclo(g)$ est la caractéristique cyclomatique de $g$ (\emph{i.e.} la longueur du plus long cycle de la plus petite base de cycles de $g$), si $g$ contient un cycle, $2$ sinon. 

En réalité, \cite{BPV04c} prouve que prendre $\alpha\geq trou(g)-2$ assure que le protocole converge en un temps fini vers une configuration de $\Gamma_1$ et que prendre $K>cyclo(g)$ assure que chaque processeur incrémente infiniment souvent son horloge. Par définition, nous savons que $trou(g)$ et $cyclo(g)$ sont majorés par $n$.

\paragraph{Protocole d'exclusion mutuelle.} L'idée principale de notre protocole est d'exécuter l'unisson auto-stabilisant de \cite{BPV04c} présenté précédemment, avec une taille d'horloge particulière et d'accorder le privilège à un processeur seulement lorsque son horloge atteint une certaine valeur. La taille de l'horloge doit être suffisante pour assurer qu'au plus un processeur soit privilégié dans toute configuration de $\Gamma_1$. Si la définition du prédicat $privilege$ garantit cette propriété, alors la stabilisation de notre protocole découle de celle de l'unisson sous-jacent.

Plus précisément, nous choisissons une horloge bornée $\mathcal{X}=(cerise(\alpha,K),\phi)$ avec $\alpha=n$ et $K=(2.n-1)(diam(g)+1)+2$ et nous définissons $privilege_v \equiv (r_v=2.n+2.diam(g).id_v)$. Notre protocole, baptisé $\mathcal{EMSS}$ (pour $\mathcal{E}$xclusion $\mathcal{M}$utuelle $\mathcal{S}$péculativement $\mathcal{S}$tabilisante) est présenté en Algorithme 1. Ce protocole est identique à celui de \cite{BPV04c} excepté pour la taille de l'horloge et la définition du prédicat $privilege$ (qui n'interfère pas avec le protocole).

\begin{algorithm}
\caption{$\mathcal{EMSS}$: Protocole d'exclusion mutuelle pour le processeur $v$}
\small
\textbf{Constantes :}\\
$\begin{array}{lll}
id_v\in ID && n\in\mathbb{N} \\
\mathcal{X}=(cerise(n,(2.n-1)(diam(g)+1)+2),\phi) &&  diam(g)\in\mathbb{N}
\end{array}$\\
\textbf{Variable :}\\
$\begin{array}{lll}
r_v\in \mathcal{X} & : & \text{registre de } v
\end{array}$\\
\textbf{Prédicats :}\\
$\begin{array}{lcl}
privilege_v \equiv (r_v=2.n+2.diam(g).id_v) &&
correct_v(u) \equiv (r_v\in stab_\mathcal{X})\wedge(r_u\in stab_\mathcal{X})\wedge(d_K(r_v,r_u)\leq 1)\\
tousCorrects_v \equiv \forall u\in vois(v), correct_v(u) &&
etapeNorm_v \equiv tousCorrects_v\wedge(\forall u\in vois(v),r_v\leq_l r_u)\\
reInit_v \equiv \neg tousCorrects_v\wedge(r_v\notin init_\mathcal{X}) &&
etapeConv_v \equiv r_v\in init^*_\mathcal{X}\wedge\forall u\in vois(v),(r_u\in init_\mathcal{X}\wedge r_v\leq_{init} r_u)\\
\end{array}$\\
\textbf{Règles :}\\
$\begin{array}{lclcl}
NA :: etapeNorm_v \longrightarrow r_v:=\phi(r_v) &&
CA :: etapeConv_v \longrightarrow r_v:=\phi(r_v) &&
RA :: reInit_v \longrightarrow r_v:=-n
\end{array}$
\normalsize
\end{algorithm}

Il est à noter que, par définition du prédicat $privilege$, deux processeurs ne peuvent pas être simultanément privilégiés dans une configuration de $\Gamma_1$ (dans laquelle l'écart entre leurs horloges est d'au plus $diam(g)$). L'auto-stabilisation du protocole d'unisson de \cite{BPV04c} permet alors de déduire le théorème suivant (dont la preuve détaillée est disponible dans \cite{DG13c}).

\begin{theorem}\label{th:correctness}
$\mathcal{EMSS}$ est un protocole auto-stabilisant pour $spec_{EM}$ sous $did$.
\end{theorem}

L'analyse du temps de stabilisation de notre protocole est disponible dans \cite{DG13c}. Pour le cas du démon synchrone, elle repose sur l'observation que, dans le pire cas, un seul processeur ré-initialise son horloge durant la première étape d'une exécution synchrone. Après cela, deux sections critiques concurrentes ne sont possibles que si cette ré-initialisation sépare deux groupes non vides de processeurs synchronisés, ce qui n'est possible que durant les $\left\lceil diam(g)/2\right\rceil$ étapes d'une une exécution synchrone (bien que la ré-initialisation puisse prendre plus longtemps pour couvrir tout le système). Pour le cas du démon inéquitable distribué, nous utilisons le fait que le temps de stabilisation de l'unisson majore celui de notre protocole.

\begin{theorem}\label{th:stabufd}
$temps\_stab(\mathcal{EMSS},ds)\leq\left\lceil diam(g)/2\right\rceil$ et $temps\_stab(\mathcal{EMSS},did)\in O(diam(g).n^3)$
\end{theorem}

Le résultat de borne inférieure suivant nous montre l'optimalité de notre protocole spéculativement stabilisant pour les exécutions synchrones (sa preuve est disponible dans \cite{DG13c}). Il repose sur l'existence d'historiques indistinguables pour tout protocole qui convergerait plus rapidement, ce qui permet de construire un contre-exemple à la stabilisation d'un tel protocole.

\begin{theorem}\label{th:lowerBound}
Tout protocole $\pi$ auto-stabilisant pour $spec_{EM}$ vérifie $temps\_stab(\mathcal{\pi},ds)\geq\left\lceil diam(g)/2\right\rceil$.
\end{theorem}

\section{Perspectives}

Cet article ouvre une nouvelle voie de recherche en auto-stabilisation en introduisant la notion de stabilisation spéculative. Nous appliquons cette notion au problème de l'exclusion mutuelle en fournissant le premier protocole spéculativement stabilisant qui soit optimal pour les exécutions synchrones. Il serait intéressant d'appliquer cette approche à d'autres problèmes fondamentaux, d'optimiser les protocoles auto-stabilisants pour différents adversaires et de fournir un outil de composition qui fournirait de manière automatique des protocoles spéculativement stabilisants.

\small
\bibliographystyle{plain}
\bibliography{biblio}

\end{document}